\documentclass[twocolumn,showpacs,preprintnumbers,
amsmath,amssymb,floatfix,nofootinbib]{revtex4-1}
\usepackage{graphicx}
\usepackage{dcolumn}
\usepackage{bm}
\usepackage{amsthm}
\usepackage{amsmath}
\usepackage{amsfonts}
\usepackage{color}

\usepackage{caption}
\usepackage{subcaption}

\newcommand{\be}{\begin{equation}}
\newcommand{\ee}{\end{equation}}
\newcommand{\ba}{\begin{eqnarray}}
\newcommand{\ea}{\end{eqnarray}}
\newcommand{\non}{\nonumber}
\newcommand{\n}[1]{\label{#1}}
\newcommand{\eq}[1]{(\ref{#1})}

\newcommand{\BM}[1]{{\mbox{\boldmath $#1$}}}

\newcommand{\vp}{v_{_{\perp}}}

\newcommand{\ce}{{\cal{E}}}
\newcommand{\cl}{{\cal{L}}}
\newcommand{\cb}{{\cal{B}}}
\newcommand{\tr}{\rho}

\newcommand{\hh}{\, ,\hspace{0.5cm}}

\begin{document}

\title{Critical escape velocity for a charged particle moving around a weakly magnetized Schwarzschild black hole}
\author{A. M. Al Zahrani}
\email{ama3@ualberta.ca}
\author{Valeri P. Frolov}
\email{vfrolov@ualberta.ca}
\author{Andrey A. Shoom}
\email{ashoom@ualberta.ca}
\affiliation{Theoretical Physics Institute, University of Alberta, Edmonton, Alberta T6G 2E1, Canada}
\begin{abstract}
We discuss charged particles motion in a spacetime of a weakly magnetized static non-rotating black hole. We study
under which conditions a charged particle originally revolving around the black hole at a circular orbit after being kicked by another particle or photon can escape to infinity. We determine the escape velocity for particles at the innermost stable circular orbits and discuss the properties of particles moving with near-critical velocity. We show that in a general case such a motion is chaotic.
\end{abstract}

\pacs{04.70.Bw, 04.25.-g, 04.70.-s, 97.60.Lf \hfill
Alberta-Thy-01-13}

\maketitle

\section{Introduction}

The mechanism of jets formation in black holes is one of the most intriguing problems of modern astrophysics. The  matter in the accretion disk and the black hole rotation might provide sufficient energy to form and support the jets. It is plausible that the key role in the transfer  mechanism of this energy to the jets is played by magnetic fields \cite{KSKM,DKBS,KN}.  Both widely discussed models, Blandford-Znajek mechanism \cite{BZ,Par} and Penrose effect for magnetic fields \cite{koide:03,KSKM}, assume that in the vicinity of  the black hole horizon there exists a sufficiently strong magnetic field. This field does not change the black hole geometry, but its interaction with charged particles and plasma is important. Such black holes are called {\em weakly magnetized}. Study of such problems in all their complexity requires 3D numerical simulations of the magnetohydrodynamics (MHD) in a strong gravitational field. (For discussion and references, see e.g.  \cite{Pun}).

Quite often when dealing with  such a complicated problem it might be instructive to consider first its different simplifications, which can be treated either analytically, or by integrating ordinary differential equations. Motion of a charged particle in a weakly magnetized black hole is an important example.
In this paper we discuss some aspects of this problem. Namely, we assume that a non-rotating black hole of mass $M$ is surrounded by a static axi-symmetric test  magnetic field, which is homogeneous at infinity, where it has the strength $B$. In the absence of the magnetic field the geodesic equations  for the particle are completely integrable. This property is also valid for the motion of the charged particle in a weakly magnetized black hole, provided such a particle moves in the equatorial plane orthogonal to the direction of the magnetic field. This case was studied in details in \cite{GP,AG,FS}. A similar problem for weakly charged rotating black holes was discussed in \cite{AO}. The main result of this study is that in the presence of the magnetic field
the innermost stable circular orbit (ISCO) of a charged particle is located close to the black hole horizon. In this sense, the action of the magnetic field on a charged particle is similar to the action of the black hole rotation on a neutral particle. In particular, weakly magnetized black holes may play a role of particle accelerators \cite{FJ,J}, similar to the fast rotating black holes (see e.g. \cite{Banados:2009pr,Banados:2010kn,Harada:2011pg,Zaslavskii:2010pw,Jacobson:2009zg,Berti:2009bk} and the references therein).

In this paper we consider motion of a charged particle in a weakly magnetized black hole out of the equatorial plane. We focus on the following problem: Suppose the particle revolving around the black hole in the equatorial plane is kicked out of it by another particle or photon. Under which conditions can such a particle leave the black hole vicinity and escape to infinity? In our model the magnetic field far from the black hole is homogeneous.
Hence, at a far distance from the black hole, where its gravitational field is weak and can be neglected, the charged particle moves in the homogeneous magnetic field. In this case, the corresponding equations are completely integrable.  However, before the particle reaches the spatial infinity, it passes through the region where both of the fields, the gravitational and magnetic, affect its motion, so that the dynamical system looses its complete integrability and the motion of the charged particle may become chaotic.

Several examples demonstrating a similar chaotic behaviour are known. For instance, even in the absence of the magnetic field chaos arises for the motion of a particle when a spherically symmetric metric of a black hole is perturbed.  For axially symmetric deformation of a black hole this was demonstrated  in \cite{Maeda}. There exist also several papers discussing a chaotic particle motion in the Majumdar-Papapertou metric, describing the spacetime with two or more extremely charged black holes in equilibrium \cite{Yurt,DFC}. In the vicinity of each of these black holes the spherically symmetric Reissner-Nordstrom metric is deformed by the common action of the other black holes. Numerical analysis of a charged particle trajectories in the presence of a toroidal magnetic field in the Schwarzschild spacetime was analyzed in \cite{PS}.  The results presented there illustrate that under special conditions charged particles moving in the vicinity of the black hole under the action of the
toroidal magnetic field can be ejected
to infinity along the axis of symmetry. Even in the absence of a black hole, when a charged particle moves in a non-uniform magnetic field it often has either trapped chaotic motion \cite{BZ1}, or chaotic scattering behavior \cite{BZ2,ABZ}. Chaotic motion of a charged particle in the Ernst spacetime representing a magnetized black hole was analysed in \cite{KV}.

In this paper we demonstrate that the charged particle motion near a weakly magnetized black hole is generically chaotic. We find the critical escape velocity for such a particle required to escape to infinity and discuss some properties of the near-critical motion.
The paper is organized as follows: In Sec. II we discuss our model and present an expression for the escape velocity for a neutral particle. In Sec. III we present the equations of motion of a charged
particle moving around a weakly magnetized Schwarzschild black hole.  Section IV discusses  scattering data for our problem. Dimensionless form of the equations and the initial conditions for the particle escape from ISCO orbits are given in Sec. V. In Sec. VI we give several examples of qualitatively different orbits of a charged particle in the weakly magnetized black hole. Basin-boundary analysis of the trajectories is applied for an analysis of the charged particle motion in Sec. VII. There we demonstrate the chaotic properties of the trajectories and determine the fractal dimensions in the proper domains. General discussion of the results is given in Sec. VIII. In this paper we use the sign conventions adopted in \cite{MTW} and units where $c=1$.

\section{Escape velocity for a neutral particle}

Before considering the escape velocity problem for a charged particle in a weakly magnetized black hole, let us
remind the well-known results for a similar problem in a simpler case when a particle is neutral and the magnetic field is absent. The background Schwarzschild metric is
\ba\n{5}
ds^2&=&-fdt^2+f^{-1}dr^2+r^2 d\omega^2\, ,\\
d\omega^2&=&d\theta^2+\sin^2\theta\,d\phi^2\hh
f=1-\frac{r_g}{r}\,.
\ea
Here $r_g=2GM$ is the gravitational radius of the black hole. There exist three commuting integrals of motion. Two of them are generated by the Killing vectors
\be\n{Kil}
\BM{\xi}_{(t)}=\xi_{(t)}^{\mu} \partial_{\mu}=\partial_t\hh \BM{\xi}_{(\phi)}=\xi_{(\phi)}^{\mu} \partial_{\mu}=\partial_{\phi}\,,
\ee
reflecting invariance with respect to time translations and rotations around the symmetry axis.
The corresponding conserved quantities are the specific energy ${\cal E}$ and the specific azimuthal angular momentum ${\cal L}_{z}$,
\ba
{\cal E} &\equiv&-p_{\mu}\xi^{\mu}_{(t)}/m=\dot{t}f\,,\\
{\cal L}_{z} &\equiv&p_{\mu}\xi^{\mu}_{(\phi)}/m=\dot{\phi}\,r^{2}\sin^{2}\theta\,.\n{L}
\ea
Here $m$ is the mass of the particle, $u^{\mu}$ and $p^{\mu}=mu^{\mu}$ are its 4-velocity and 4-momentum, respectively. Here and in what follows, the overdot denotes the derivative with respect to the proper time. The third integral of motion is the square of the specific total angular momentum
\be\n{am}
\cl^{2}\equiv r^{4}\dot{\theta}^{2}+\frac{\cl_{z}^{2}}{\sin^{2}\theta}=r^{2}\vp^{2}+\frac{\cl_{z}^{2}}{\sin^{2}\theta}\, .
\ee
Here we denoted by  $\vp$ the following quantity:
\be\n{vp}
\vp\equiv -r\dot{\theta}_{o}\, .
\ee
Using the normalization condition $\BM{u}^2=-1$ one obtains
\be
\dot{r}^{2}=\ce^{2}-U\hh U=f(1+\cl^{2}/r^{2})\,.
\ee

The motion of the particle is planar. Let this plane coincides with the equatorial plane. Then $\dot{\theta}=0$ and the effective potential for the radial motion takes the form
\be
U=\tilde{U}\equiv f(1+\cl^{2}_z/r^{2})\,.
\ee
Consider a particle at the circular orbit $r=r_o$, where $r_o$ is the local minimum of the effective potential $\tilde{U}$. This orbit exists for $r_o\in(3r_{g},\infty)$.
The corresponding specific energy and azimuthal angular momentum are
\be
\ce_{o}=\frac{\sqrt{2}(r_{o}-r_{g})}{\sqrt{r_{o}(2r_{o}-3r_{g})}}\hh |\cl_{zo}|=\frac{r_{o}\sqrt{r_{g}}}{\sqrt{2r_{o}-3r_{g}}}\,.
\ee
The ISCO  is defined by $r_{o}=3r_{g}$, which corresponds to a convolution point of the effective potential. For ISCO we have $\ce_{\text{ISCO}}=2\sqrt{2}/3\approx0.943$ and $|\cl_{z\text{ISCO}}|=r_{g}\sqrt{3}$.

Suppose now that the particle at a circular orbit collides with another particle or photon, so that
after the collision it will move  within a new plane tilted with respect to the original equatorial plane. In a general case, all three types of its motion are possible: (i) bounded motion, (ii) escape to infinity, and (iii) capture by the black hole. The result certainly depends on the details of the collision mechanism.
For small values of the transferred energy and momentum the orbit will be only slightly perturbed.  However, for larger values of $\ce-\ce_{o}$ the particle can go away from the initial plane and finally can be captured by the black hole or escape to infinity.

In a general case, as a result of the collision, the particle will have new integrals of motion: $\ce$, $\cl_z$ and $\cl^2$. For the case of a neutral particle in the Schwarzschild black hole one can easily obtain the conditions of escape in an analytical form. To be able to obtain results which allow rather simple analysis and presentation we simplify the problem and reduce the space of initial data to a one-parameter set by imposing the following restrictions: (i) the azimuthal angular momentum is not changed, and (ii) the initial radial velocity after the collision remains the same, $\dot{r}_o=0$. Under these restrictions there exists only one parameter which determines the motion of the particle, namely the new value of its energy.
Under these conditions,  as a result of the collision, the particle acquires a velocity $\vp$ in the direction orthogonal to the equatorial plane [see \eq{vp}].

After the collision, the total angular momentum and the energy of the particle are
\ba
\cl^{2}&=&r_o^{2}\vp^{2}+\cl_{z}^{2}\,,\n{L2}\\
\ce&=&\sqrt{\ce_{o}^{2}+\vp^{2}(r_{o}-r_{g})/r_{o}}\,.
\ea
During the particle motion its polar coordinate $\theta$ changes within the interval $[\arcsin(|\cl_{z}|/\cl),\pi-\arcsin(|\cl_{z}|/\cl)]$.

As a result of the collision, the total angular momentum of the particle changes from its original value $\cl_{z}^{2}$ to $\cl^{2}$ given by Eq. \eq{L2}. The effective potential $U$ defined by the new value of $\cl^{2}$ is greater than that before the collision and its extrema correspond to $r=r_{\pm}$, where
\be
r_{\pm}=\frac{\cl^{2}\pm\sqrt{\cl^{2}(\cl^{2}-3r_{g}^{2})}}{r_{g}}\,,
\ee
and $r_{-}$ ($r_{+}$) is a monotonically decreasing (increasing) function of $\cl^{2}$. Thus, immediately after the collision the particle is still at the turning point ($\dot{r}_o=0$), which is located between the maximum and minimum of the effective potential $U$. Therefore, the particle will escape to infinity if $\ce\geq1$, or
\be
|\vp|\geq\sqrt{\frac{r_{o}(1-\ce_{o}^{2})}{(r_{o}-r_{g})}}\,.
\ee
In particular, for ISCO we have the escape condition $|\vp|\ge \vp^{esc}\geq 1/\sqrt{6}$, where the last equality corresponds to $\ce=1$.

The black hole metric \eq{5} has the evident discrete symmetries
\be
\phi\to 2\pi-\phi \mbox{   and   }\theta\to \pi-\theta\, .
\ee
These symmetries imply the symmetry of the problem with respect to the following transformation:
\be
\cl_z\to -\cl_z \mbox{   and   }\vp\to -\vp\,.
\ee

\section{Charged particles in a magnetized black hole}

We consider now the case of weakly magnetized black holes.
We  assume that a particle has the electric charge $q$ and that its motion is affected by the magnetic field in the black hole exterior.
Namely, we assume that there exists a magnetic field in the black hole vicinity which
is static, axi-symmetric and homogeneous at the spatial infinity where it has the strength $B$.
According to the procedure given in \cite{Wald,AG}, such a magnetic field can be constructed as follows:
Because the metric \eq{5} is Ricci flat, the Killing vectors \eq{Kil} obey the equation
\be\n{6}
\xi^{\mu;\nu}_{\,\,\,\,\,\,\,;\nu}=0\, .
\ee
This equation coincides with the Maxwell equation for a  4-potential $A^\mu$ in the Lorenz gauge
$A^\mu_{\,\,\,;\mu}=0$.  The special choice
\be\n{A}
A^{\mu}={B\over 2}\xi^{\mu}_{(\phi)}\, ,
\ee
corresponds to a test magnetic field, which is homogeneous at the spatial
infinity where it has the strength $B$. The electric 4-potential
\eq{A} is invariant with respect to the isometries corresponding to the Killing vectors, i.e.,
\be
({\cal L}_{{\bm \xi}}A)_{\mu}=A_{\mu,\nu}\xi^{\nu}+A_{\nu}\xi^{\nu}_{\,\,,\mu}=0\,.
\ee
A magnetic field is defined with respect to an observer whose 4-velocity is $u^{\mu}$ as follows:
\be\n{B}
B^{\mu}=-\frac{1}{2}e^{\mu\nu\lambda\sigma}F_{\lambda\sigma}u_{\nu}\,,
\ee
where
\be
e^{\mu\nu\lambda\sigma}=\frac{\epsilon^{\mu\nu\lambda\sigma}}{\sqrt{-g}}\hh \epsilon_{0123}=+1\hh g=\text{det}(g_{\mu\nu})\,,
\ee
and
\be
F_{\mu\nu}=A_{\nu,\mu}-A_{\mu,\nu}\,.
\ee
For a local observer at rest in the Schwarzschild spacetime \eq{5} we have $u_{o}^{\mu}=f^{-1/2}\xi^{\mu}_{(t)}$, and Eqs. \eq{A} and \eq{B} give
\be\n{Bo}
B^{\mu}=Bf^{1/2}\left(\cos\theta\delta^{\mu}_{r}-\frac{1}{r}\sin\theta\delta^{\mu}_{\theta}\right)\,.
\ee
At the spatial infinity the magnetic field is directed along the vertical $z$ axis. In what follows, we assume that the field is directed upward, therefore, we shall take $B>0$.

In a curved spacetime a charged particle of mass $m$ moving in an external electromagnetic field $F_{\mu\nu}$ obeys the equation
\be
m\frac{D{u}^{\mu}}{d\tau}=qF^{\mu}_{\,\,\,\nu}\,u^{\nu}\,,\n{eq1}
\ee
where $D/d\tau$ is the covariant derivative defined with respect to the metric \eq{5}, $u^{\mu}=\dot{x}^{\mu}$ is the particle 4-velocity, $u^{\mu}u_{\mu}=-1$, and
$q$ is its charge.

Denote by $P_{\mu}\equiv m u_{\mu}+qA_{\mu}$ the generalized 4-momentum of the particle. Then the conserved quantities corresponding to the symmetries of the problem, the specific energy and azimuthal angular momentum, are defined as follows:
\ba
\ce&\equiv&-\xi^{\mu}_{(t)}P_{\mu}/m=\dot{t}f\,,\n{8}\\
\cl_{z}&\equiv&\xi^{\mu}_{(\phi)}P_{\mu}/m=\left(\dot{\phi} +\frac{qB}{2m}\right)r^2\sin^2\theta\,.\n{9}
\ea
Let us denote
\be\n{not}
\cb\equiv\frac{qB}{2m}\,.
\ee

Using these conserved quantities and the normalization of the 4-velocity vector one gets
\be
\dot{t}=f^{-1}\ce\hh \dot{\phi}={ \cl_{z}\over r^2 \sin^2\theta}-\cb\,.\n{t}
\ee
The $\rho$ and $\theta$ components of Eq. (\ref{eq1}) give, respectively
\begin{eqnarray}
\ddot{r}&=&\frac{1}{2}(2r-3r_{g})\left(\dot{\theta}^2+\frac{\cl_{z}^{2}}{r^{4}\sin^{2}\theta}\right)+\frac{r_{g}(2\cl_{z}\cb-1)}{2r^{2}}\non \\
&-&\frac{\cb^{2}}{2}(2r-r_{g})\sin^{2}\theta, \n{r}\\
\ddot{\theta}&=&-\frac{2}{r}\dot{r}\dot{\theta}+\frac{\cl_{z}^2\cos\theta}{r^{4}\sin^{3}\theta}
-\cb^{2}\sin\theta\cos\theta\,.\n{th}
\end{eqnarray}
The normalization condition $\BM{u}^2=-1$ gives the following first order equation
\ba
&&{\ce}^2=\dot{r}^2+r^2f\dot{\theta}^2+U_{\text{eff}}\,,\n{e}\\
&&U_{\text{eff}}=f\left[1+ r^2\sin^2\theta\left(\frac{\cl_{z}}{r^{2}\sin^{2}\theta}-\cb\right)^{2}\ \right]\,.\n{ue}
\ea
This equation is a constraint. If it is satisfied at the initial moment of time, then it is always valid, provided the dynamics of $r(\tau)$ and $\theta(\tau)$ is controlled by Eqs.  (\ref{r}) and (\ref{th}).

Let us discuss the symmetry properties of Eqs. \eq{t}--\eq{ue}.
First of all, these equations  are invariant under the transformations
\be
\phi\to-\phi\hh \cl_{z}\to-\cl_{z}\hh \cb\to-\cb\,.\n{tr}
\ee
Therefore, without the loss of generality, we can consider the particle of the positive electric charge (and hence $\cb>0$). To consider a particle of a negative charge one should apply the transformation \eq{tr}. In other words, a trajectory of a  negative charge is related to the positive charge trajectory by the transformation $\cl_{z}\to-\cl_{z}$, $\phi\to-\phi$.
However, after making the choice $\cb>0$, one needs to study both cases when $\cl_{z}$ is positive and negative. They are physically different: The change of the sign of $\cl_{z}$ corresponds to the change of the direction of the Lorentz force acting on the particle.

The system \eq{t}--\eq{ue} is also invariant with respect to the reflection $\theta\to\pi-\theta$.  This transformation preserves the initial position of the particle and changes $\vp\to-\vp$ [see Eq. \eq{vp}]. Therefore, it is sufficient to consider only positive values of the kick velocity $\vp$.

\section{Flat spacetime limit and the scattering data}

Suppose a kicked particle  escapes to infinity. Let us discuss first the asymptotic properties of such escaped particle. At the asymptotic infinity the gravitational field of the black hole vanishes. Hence, such a charged particle is moving in a practically flat spacetime with homogeneous magnetic field. This type of motion is well-known (see, e.g., \cite{Lan}). The corresponding equations can be obtained from \eq{t}--\eq{ue} by  introducing the cylindrical coordinates
\be
R=r\sin\theta\hh z=r\cos\theta\,,
\ee
and taking the limit $|z|\to\infty$ while keeping $R$ finite. The asymptotic form of the equations is
\ba
&&\dot{t}=\ce\hh \dot{\phi}={ \cl_{z}\over R^2}-\cb\,,\n{ti}\\
&&\ddot{z}=0\hh \ddot{R}=\frac{\cl_{z}^{2}}{R^{3}}-\cb^{2}R\,,\n{zr}\\
&&{\ce}^2=1+\dot{z}^2+\dot{R}^2+R^{2}\left({\cl_{z}\over R^{2}}-\cb\right)^{2}\,.\n{ui}
\ea

A solution to these equations is well-known, it represents a helix whose axis is directed along the magnetic field ${\BM B}$, i.e. parallel to the $z$ axis. If the $z$ component of the particle velocity vanishes, its trajectory becomes a circle located in a $z=const$ plane.
Let $R_{c}$ be the radius of the circle, then the particle velocity ${\BM v}$ is given by
\be
{\BM v}=-\frac{q}{m}[{\BM B}\times {\BM R}_{c}]\,.
\ee
Here $[{\BM a}\times {\BM b}]$ is a vector product of the 3-vectors ${\BM a}$ and ${\BM b}$ defined in a Euclidean space in the standard way.  For the uniform magnetic field ${\BM B}$ defined by \eq{A} we have
\be
{\BM A}=\frac{1}{2}[{\BM B}\times{\BM  R}]\,.
\ee
Therefore, the generalized 3-momentum vector of the particle ${\BM P}$ reads
\be
{\BM P}=-q[{\BM B}\times {\BM  R}_{c}]+\frac{q}{2}[{\BM B}\times{\BM  R}]\,,
\ee
and the corresponding angular 3-momentum vector about the point of intersection of the $z$-axis and the $z=const$ plane is
\be
{\BM L}=[{\BM  R}\times{\BM P}]\,.
\ee
It is directed along the $z$ axis, i.e. ${\BM L}=L{\BM n}_{z}$, where ${\BM n}_{z}$ is a unit vector which defines positive direction of $z$. If the center of the circle is located on the $z$ axis we have ${\BM  R}={\BM  R}_{c}$ and the $z$ component of the angular 3-momentum vector ${\BM L}_{c}$ reads
\be
L_{c}=-\frac{q}{2}B R_{c}^{2}<0\,.
\ee
In our notations \eq{not} we have
\be
\cl_{c}=- R_{c}^{2}\cb\hh  R_{c}=\sqrt{|\cl_{c}|/\cb}\,,
\ee
Thus, for $\dot{z}=0$ the solution to Eqs. \eq{ti}--\eq{ui} is
\be\n{rr}
 R= R_{c}\hh \phi=\phi_{0}-\frac{2\cb}{\ce}t\,,
\ee
where $\phi_{0}$ is a constant corresponding to $t=0$. The general solution to Eqs. \eq{ti}--\eq{ui} can be obtained by coordinate transformation of \eq{rr} by moving the center of the circle to another point ${\bf R}_h=( R_{h},\phi_{h})$ on the plane $z=const$ and boosting the solution in the $z$ direction. Accordingly, we derive
\ba
&& R=\sqrt{ R_{h}^{2}+ R_{c}^{2}+2 R_{h} R_{c}\cos\left(\phi_{0}-\frac{2\cb}{\ce}t-\phi_{h}\right)}\,,\\
&&\phi=\phi_{h}+\arccos\left(\frac{ R^{2}+ R_{h}^{2}- R_{c}^{2}}{2 R_{h} R}\right)\hh z=z_{0}+\frac{v_{\perp\infty}}{\ce}t\,,\non\\
\ea
where $z_{0}$ is a constant corresponding to $t=0$ and $v_{\perp\infty}$ is constant velocity along the $z$ direction.  As a result of this transformation, the azimuthal angular 3-momentum $\cl_{z}$ reads
\be\n{aam}
\cl_{z}=\cb( R_{h}^{2}- R_{c}^{2})=\cl_{c}+\cb R_{h}^{2}\,,
\ee
where $ R_{h}$ is the distance from the $z$ axis to the axis of the helix. Thus, one can see that for $\cl_{z}>0$ ($\cl_{z}<0$) the $z$ axis is located outside (inside) the circle, while for $\cl_{z}=0$ it passes through the circle.

The energy of the particle does not depend on the location of the circle and can be expressed as follows:
\be
\ce^{2}=1+\dot{z}^{2}+4\cb^{2} R_{c}^{2}=1+\dot{z}^{2}+4\cb^{2} R_{h}^{2}-4\cb\cl_{z}\,.
\ee

If the particle is at rest, we have $\ce=1$. This corresponds to $\dot{z}=0$ and $ R_{c}=0$. According to Eq. \eq{aam}, the last equality implies
\be
\cl_{z}=\cb R_{h}^{2}\geq0\,.
\ee
Here the equality sign corresponds to $ R_{h}=0$, i.e., the particle is located on the $z$ axis. If $\dot{z}=0$ and $\cl_{z}<0$, we have $ R_{c}> R_{h}\geq0$, and
\be
\ce^{2}=1+4\cb^{2} R_{c}^{2}=1+4\cb^{2} R_{h}^{2}-4\cb\cl_{z}>1\,.
\ee

Returning back to our main problem of the charged particle escape, we can formulate the corresponding scattering data as the following set: $\{\dot{z}_{\infty}, {\bf R}_{h}\}$.

\section{Dimensionless form of the equations}

After these remarks we return to our problem. We shall integrate numerically the dynamical equations. For this purpose we introduce the following dimensionless quantities:
\be
\sigma=\frac{\tau}{r_{g}}\hh \tr=\frac{r}{r_{g}}\hh \ell=\frac{\cl_{z}}{r_{g}}\hh b=\cb r_{g}\,.
\ee
The $\tr$ and $\theta$ components of the dynamical equation (\ref{eq1}) together with the expression for the energy $\ce$ take the form
\ba
\frac{d^{2}\tr}{d\sigma^{2}}&=&\frac{1}{2}(2\tr-3)\left(\frac{d\theta}{d\sigma}\right)^2 +\frac{(2\ell b-1)}{2\tr^{2}}+\frac{\ell^{2}(2\tr-3)}{2\tr^{4}\sin^{2}\theta}\non \\
&-&\frac{b^{2}}{2}(2\tr-1)\sin^{2}\theta, \n{rho}\\
\frac{d^{2}\theta}{d\sigma^{2}}&=&-\frac{2}{\tr}\frac{d\tr}{d\sigma}\frac{d\theta}{d\sigma}+\frac{\ell^2\cos\theta}{\tr^{4}\sin^{3}\theta}-b^{2}\sin\theta\cos\theta\,.\n{the}\\
{\ce}^2&=&\left(\frac{d\tr}{d\sigma}\right)^2+\tr(\tr-1)\left(\frac{d\theta}{d\sigma}\right)^{2}+U_{\text{eff}}\,,\n{ee}\\
U_{\text{eff}}&=&\left(1-\frac{1}{\tr}\right)\left[1+ {(\ell-b \tr^2\sin^2\theta)^{2}\over \tr^2\sin^2\theta} \right]\,.\n{uu}
\ea

The  energy of a particle revolving around the black hole in a circular orbit of radius $\tr_{o}$ at the equatorial plane $\theta=\pi/2$  is
\be
\ce_{o}=\left(1-\frac{1}{\tr_{o}}\right)^{1/2}\left[1+ {(\ell-b \tr_{o}^2)^{2}\over \tr_{o}^2} \right]^{1/2}\,.\n{e0}
\ee
As we did before, in the case of a neutral particle, we assume  that a kick does not change the particle's azimuthal angular momentum $\ell$, but only gives the particle the transverse velocity $\vp>0$. As a result, the particle energy changes from $\ce_{o}$ to
\be\n{enk}
\ce=\left[\ce_{o}^{2}+\frac{(\tr_{o}-1)}{\tr_{o}}\vp^{2}\right]^{1/2}\,.
\ee
We would like to know whether the particle will escape to the asymptotic infinity. To simplify the problem, we shall consider a particle initially moving in an ISCO. In this case, the parameters $\ell$ and $b$ are defined by the radius of the orbit $\tr_{o}$ as follows (see \cite{FS}):
\ba
&&\ell=\pm\frac{\tr_{o}(3\tr_{o}-1)^{1/2}}{\sqrt{2}\left(4\tr_{o}^2-9\tr_{o}+
3\pm\sqrt{(3\tr_{o}-1)(3-\tr_{o})}\right)^{1/2}}\,,\non\\
&&b=\frac{\sqrt{2}(3-\tr_{o})^{1/2}}{2\tr_{o}\left(4\tr_{o}^2
-9\tr_{o}+3\pm\sqrt{(3\tr_{o}-1)(3-\tr_{o})}\right)^{1/2}}\,.\non\\
\n{lb}
\ea
Here for $\ell>0$ we have $\tr_{o}\in(1,3]$, and for $\ell<0$ we have $\tr_{o}\in[(5+\sqrt{13})/4,3]$. For $\tr_{o}=3$ we have $\ell=\pm\sqrt{3}$ and $b=0$.
For such parametrization the magnetic field, $b$, as well as the specific angular momentum $\ell$ are uniquely specified by the radius of ISCO $\tr_{o}$, while the only one left parameter $\ce$ serves to specify the kick.

\section{Types of the trajectories}

\begin{figure*}[ht]
\begin{center}
\ba
&&\hspace{1cm}\includegraphics[width=5cm]{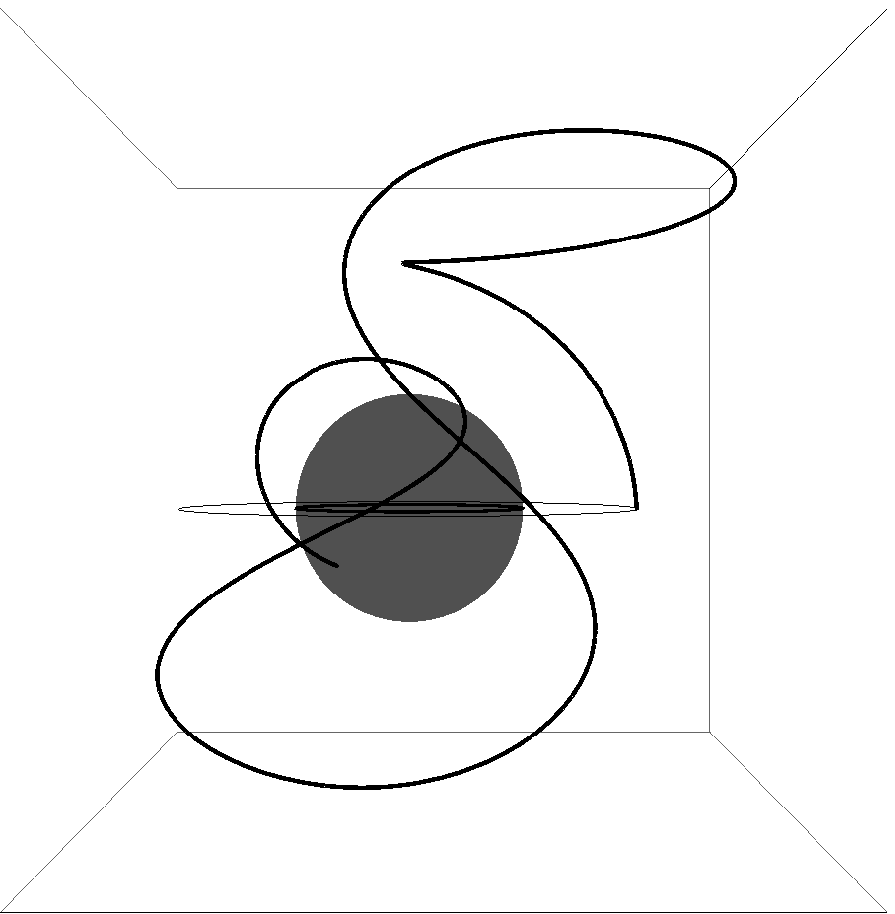}
\hspace{1cm}\includegraphics[width=5cm]{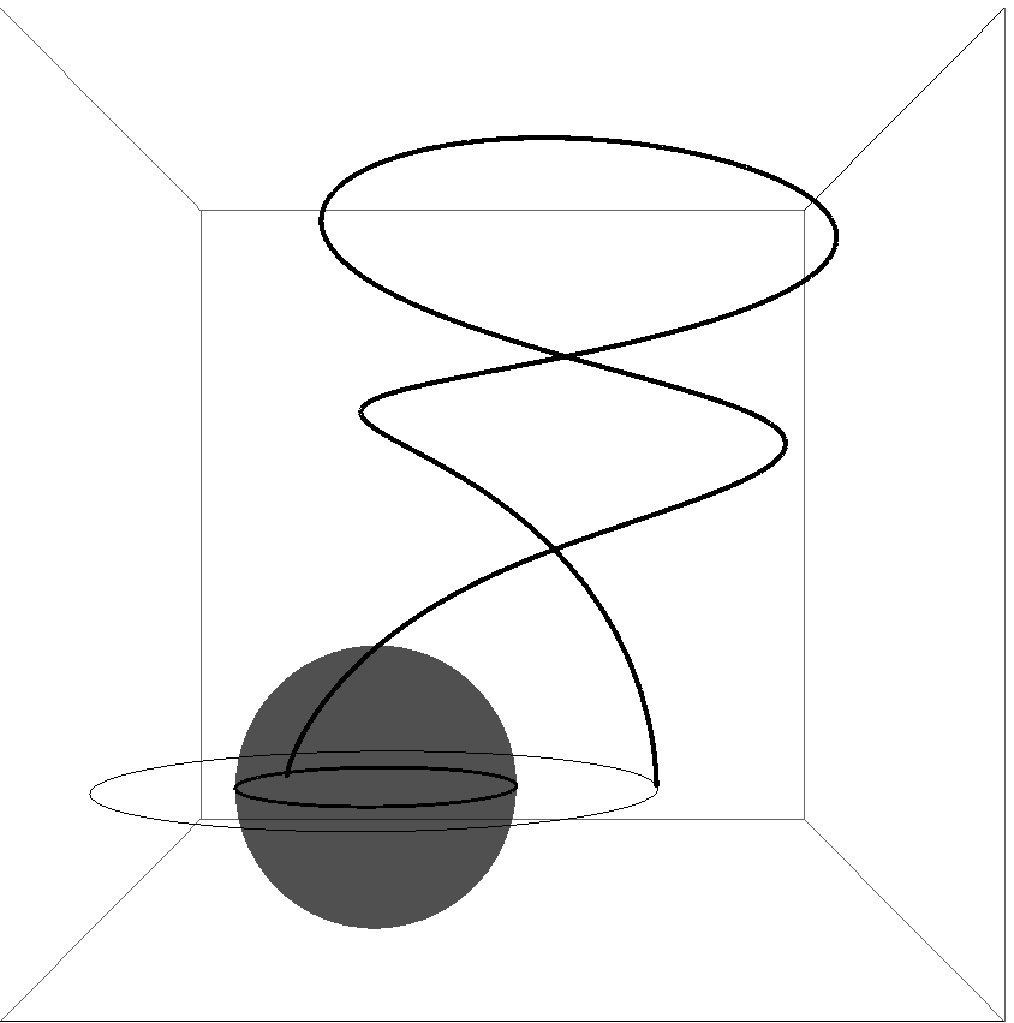}
\hspace{1cm}\includegraphics[width=5cm]{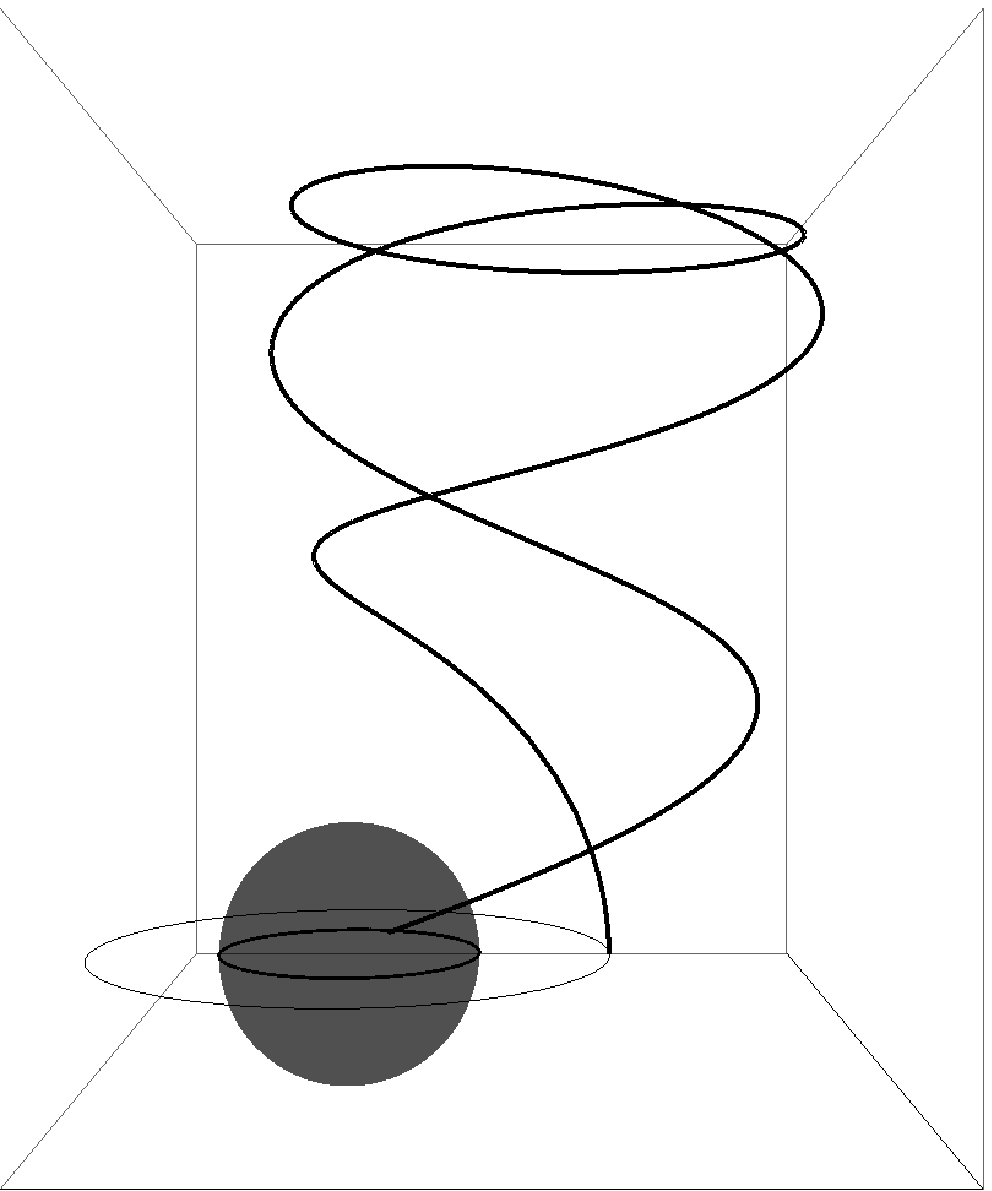}\non\\
&&\hspace{1.0cm}({\bf a})\hspace{7.8cm}({\bf b})\hspace{6cm} ({\bf c})\non
\ea
\caption{Examples of ``capture'' trajectories. In each case a charged particle is kicked up from its original ISCO at $\rho_o=2$ for $\ell>0$. The energy $\ce$ after the kick is ({\bf a}) 1.12 ({\bf b}) 1.2 and ({\bf c}) 1.3.}\label{F1a}
\end{center}
\end{figure*}
\begin{figure*}[ht]
\begin{center}
\ba
&&\hspace{1cm}\includegraphics[width=5cm]{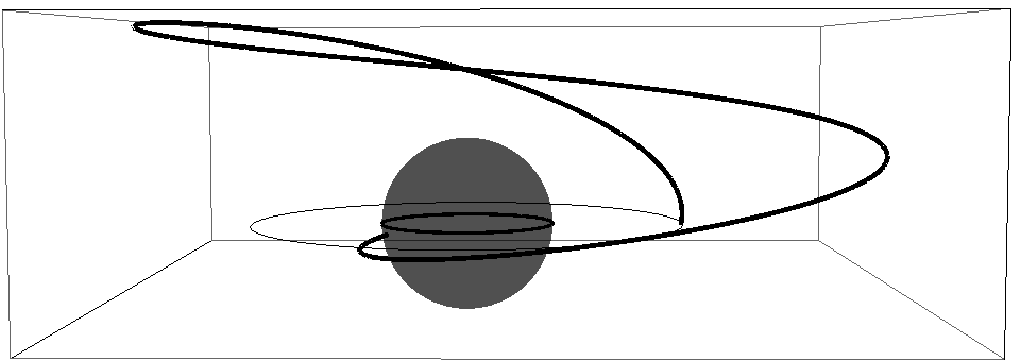}
\hspace{1cm}\includegraphics[width=5cm]{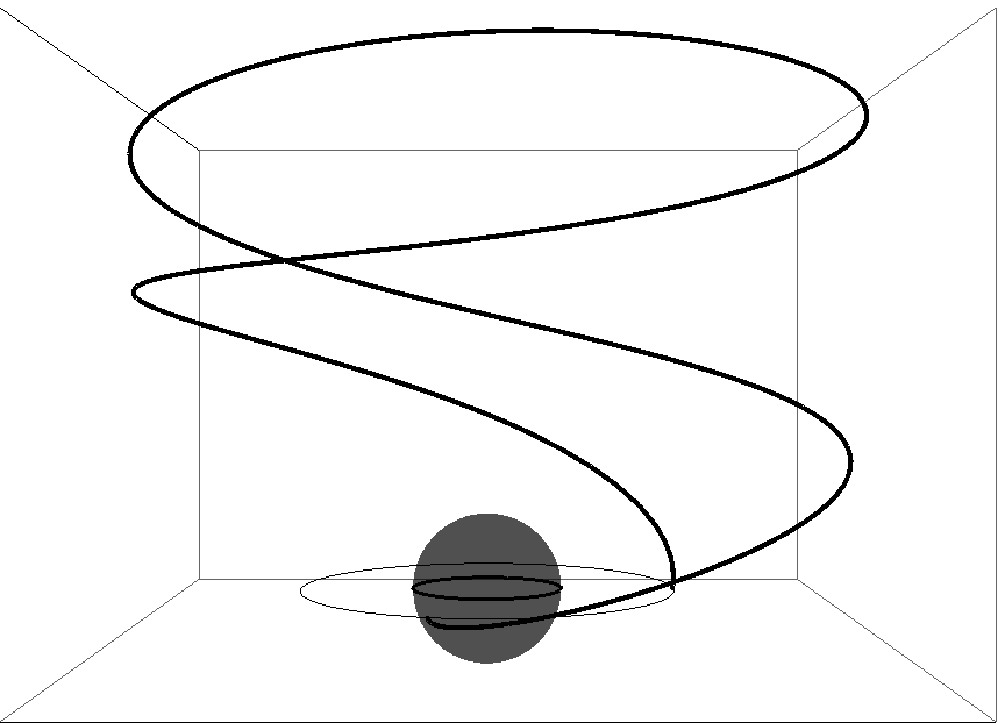}
\hspace{1cm}\includegraphics[width=5cm]{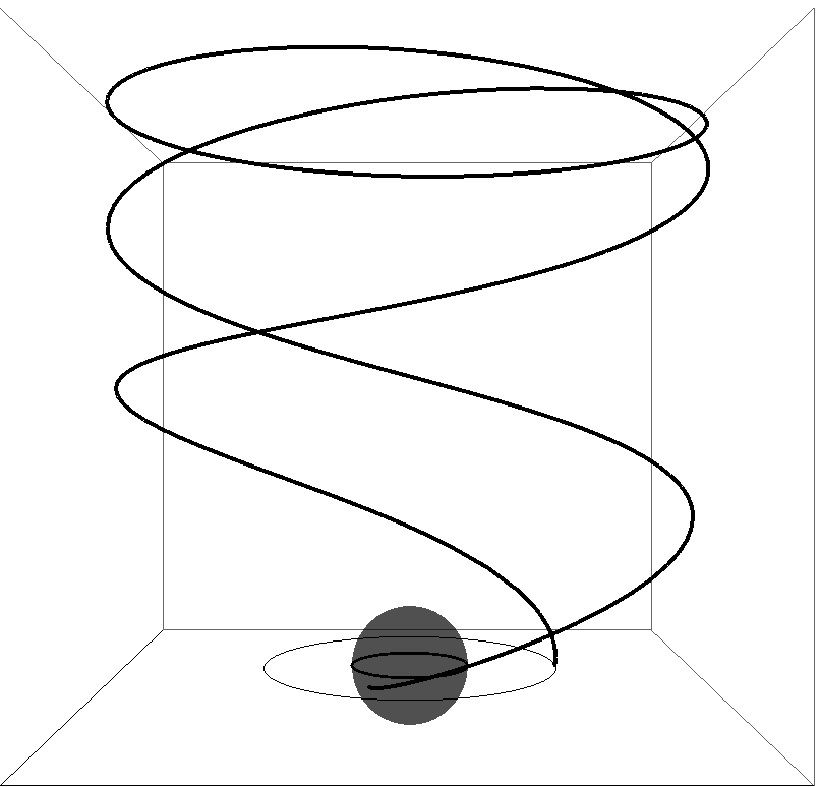}\non\\
&&\hspace{1.0cm}({\bf a})\hspace{7.8cm}({\bf b})\hspace{6cm} ({\bf c})\non
\ea
\caption{Examples of ``capture'' trajectories. In each case a charged particle is kicked up from its original ISCO at $\rho_o=2.5$ for $\ell<0$. The energy $\ce$ after the kick is ({\bf a}) 1.35 ({\bf b}) 1.475 and ({\bf c}) 1.525.}\label{F1b}
\end{center}
\end{figure*}

\begin{figure*}[ht]
\begin{center}
\ba
&&\hspace{1cm}\includegraphics[height=8cm]{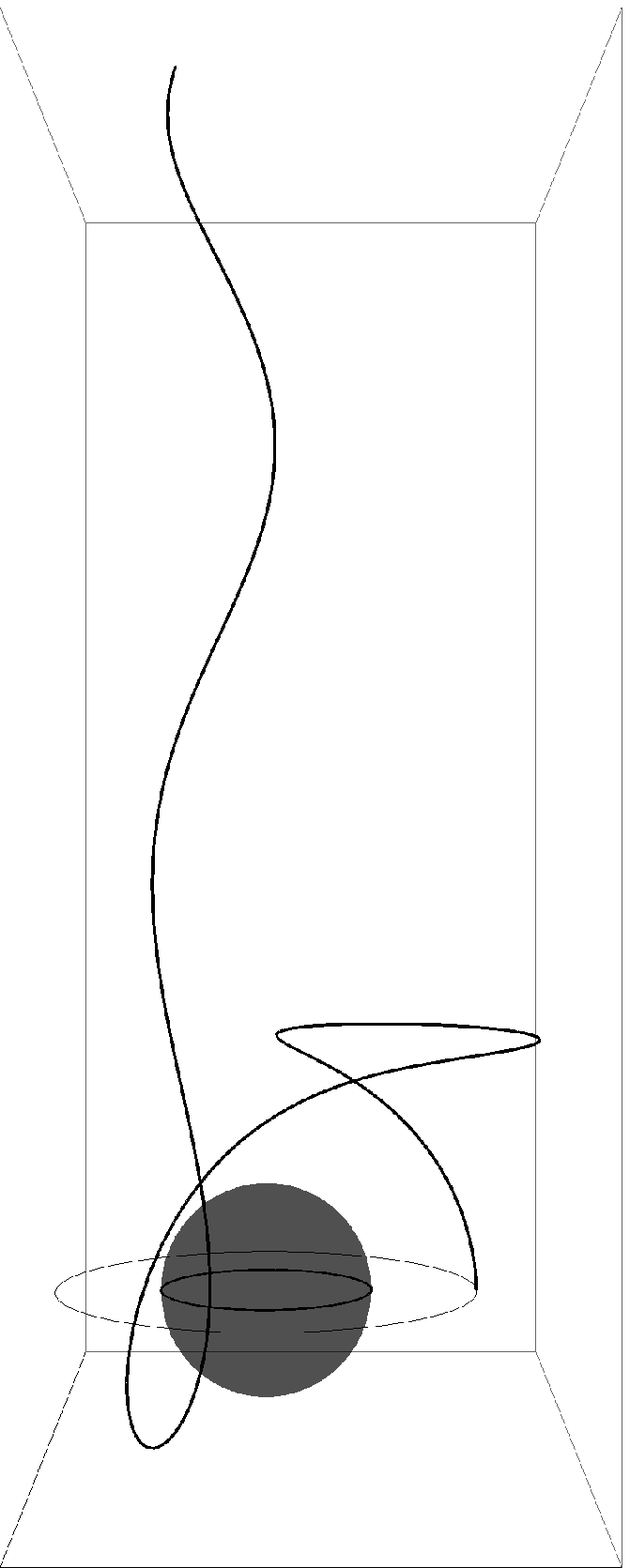}
\hspace{1cm}\includegraphics[height=8cm]{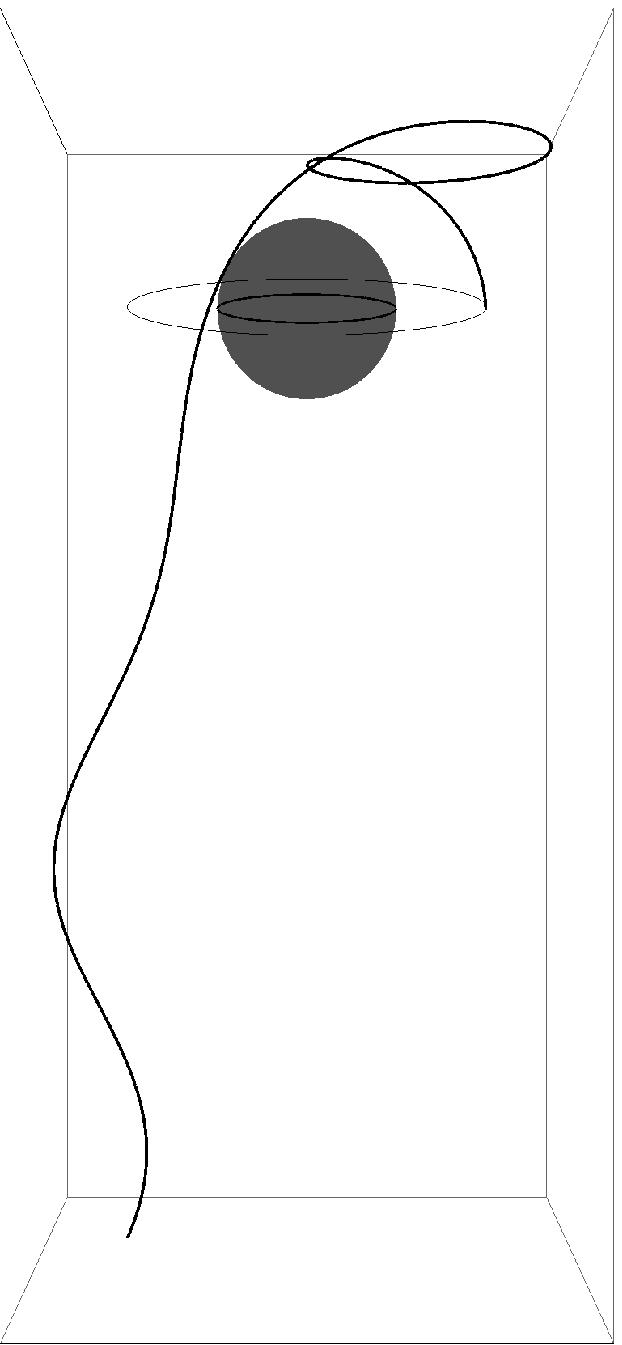}
\hspace{1cm}\includegraphics[height=8cm]{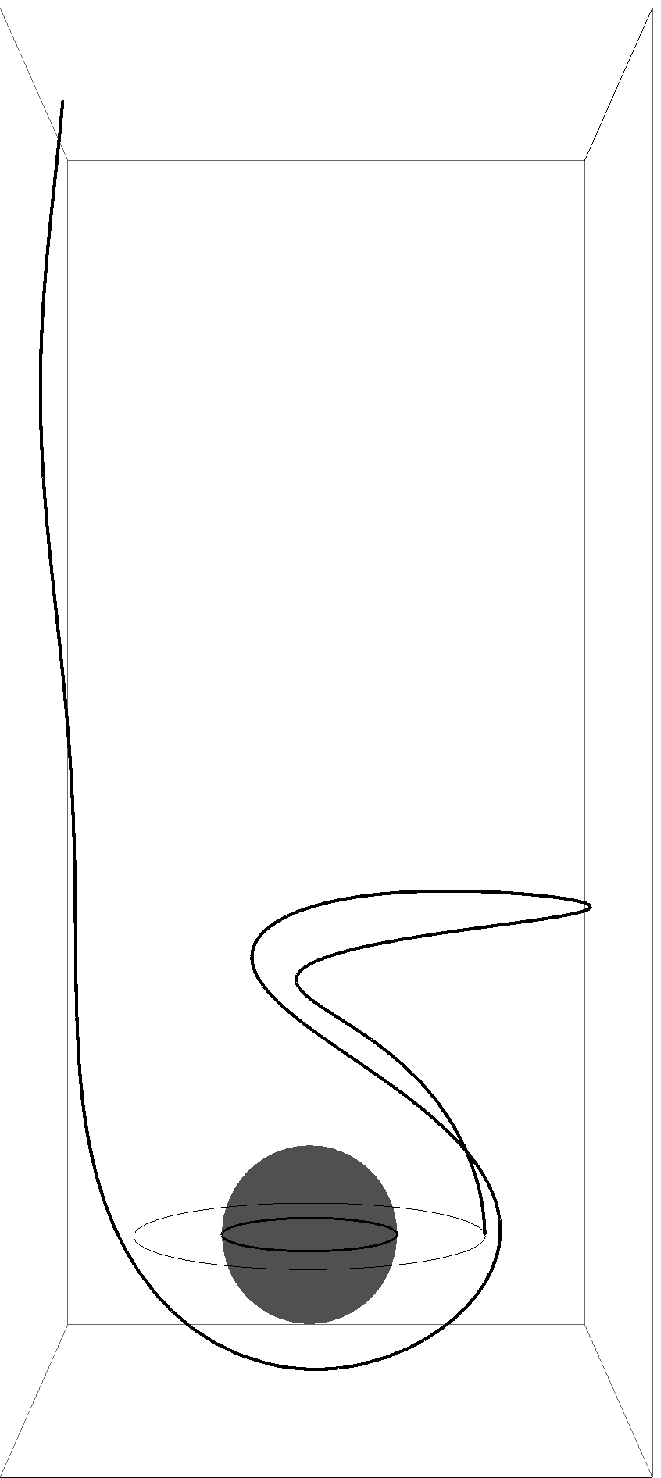}\non\\
&&\hspace{1.0cm}({\bf a})\hspace{4cm}({\bf b})\hspace{4cm} ({\bf c})\non
\ea
\caption{Examples of ``escape'' trajectories. In each case a charged particle is kicked up from its original ISCO at $\rho_o=2$ for $\ell>0$. The energy $\ce$ after the kick is ({\bf a}) 1.025 ({\bf b}) 1.05 and ({\bf c}) 1.135.}\label{F1c}
\end{center}
\end{figure*}
\begin{figure*}[ht]
\begin{center}
\ba
&&\hspace{1cm}\includegraphics[height=8cm]{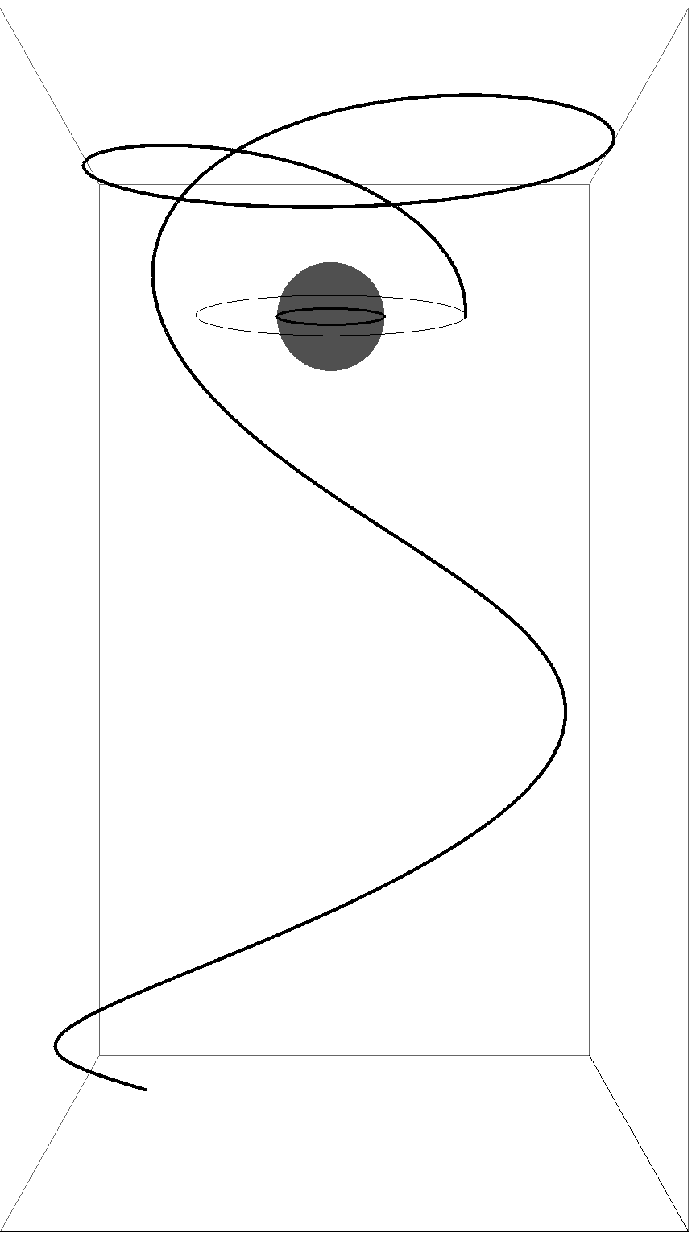}
\hspace{1cm}\includegraphics[height=8cm]{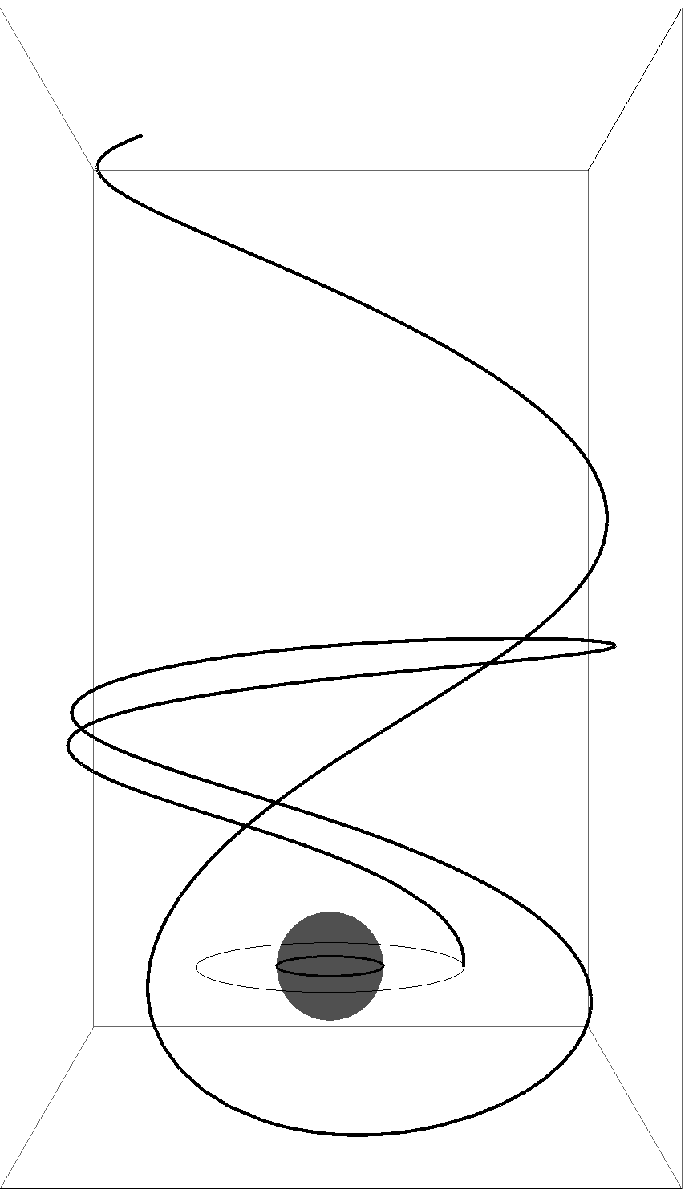}
\hspace{1cm}\includegraphics[height=8cm]{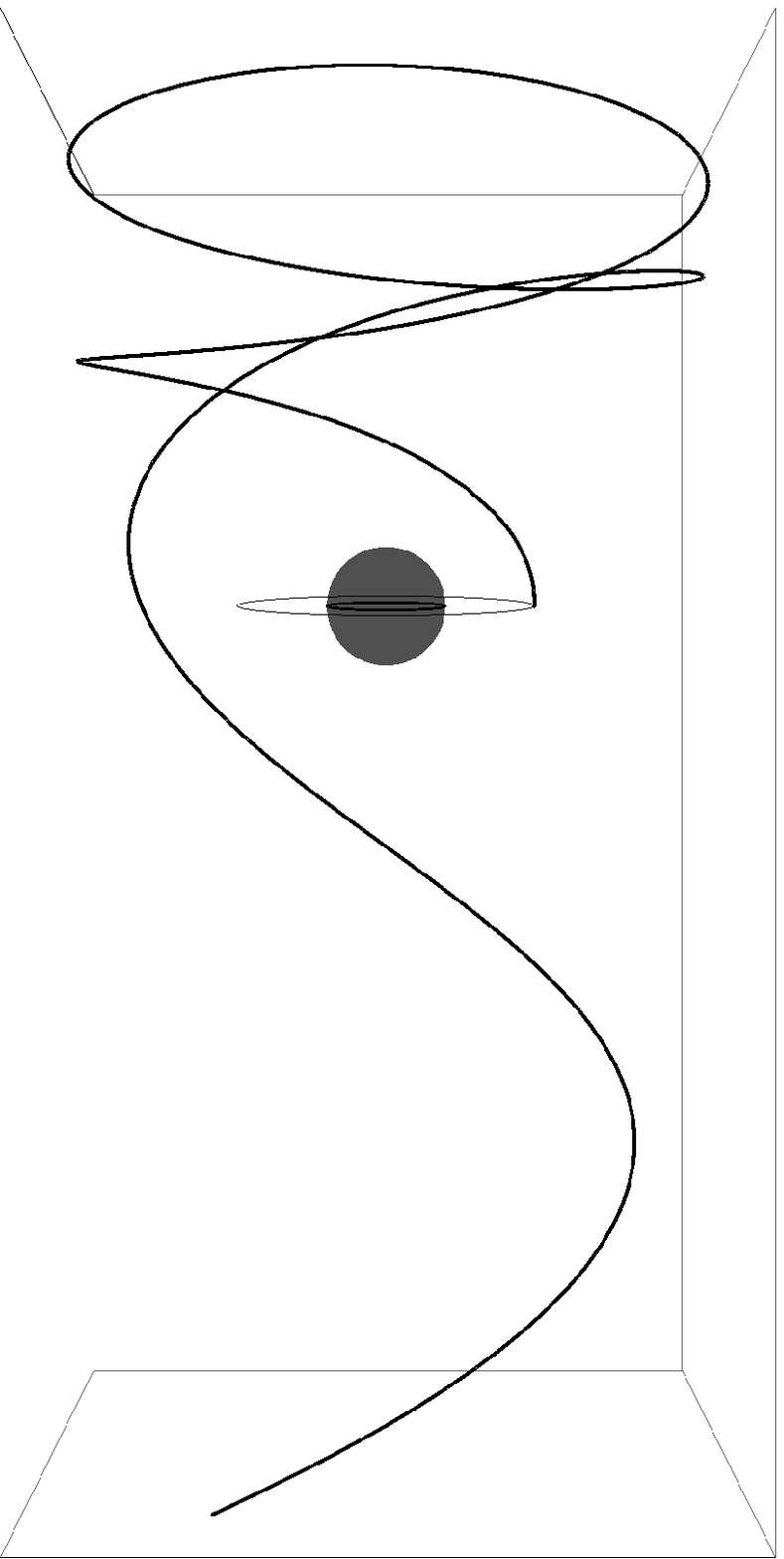}\non\\
&&\hspace{1.0cm}({\bf a})\hspace{5cm}({\bf b})\hspace{5cm} ({\bf c})\non
\ea
\caption{Examples of ``escape'' trajectories. In each case a charged particle is kicked up from its original ISCO at $\rho_o=2.5$ for $\ell<0$. The energy $\ce$ after the kick is ({\bf a}) 1.415 ({\bf b}) 1.46 and ({\bf c}) 1.5.}\label{F1d}
\end{center}
\end{figure*}

Given the orbit radius $\tr=\tr_{o}$ and the initial energy $\ce>\ce_{o}$ of the particle after the kick, we integrate the dynamical equations \eq{rho}--\eq{the} numerically. As a result of the integration, we can find a trajectory corresponding to the given initial conditions. The dynamical equations were solved using the built-in {\it Mathematica} 8.0 function \verb"NDSolve". The integral of motion of the system $\ce$ [see Eqs. \eq{ee} and \eq{enk}] was used to estimate the accuracy of the numerical solver. For our calculations the energy error was found to be less than $10^{-6}$.

Results of the numerical integration show that there are three different types of the final particle motion:
\begin{enumerate}
\item The particle is captured by the black hole.
\item The particle escapes to $z\to-\infty$.
\item The particle escapes to $z\to+\infty$.
\end{enumerate}

The outcome of the motion is considered a capture when $\rho$ reaches 1. It is considered an escape if $|z|$ reaches $10^3$. The maximum computation time was chosen $\sigma=10^5$. In escape cases, it was found that the cumulative error can reach $10^{-2}$. The accuracy of the numerical solver can be increased to achieve much better accuracy. While increasing the accuracy is not a problem when few trajectories are to be generated, it can increase the computation time greatly when the equations of motions are intergraded hundreds of thousands of times as in the case of producing basin-boundary plots (see below). However, at least in the cases we have studied, increasing the accuracy of the numerical solver does not change the final state of the motion significantly.

\begin{figure*}[ht]
\begin{center}
\ba
&&\hspace{1cm}\includegraphics[width=7cm]{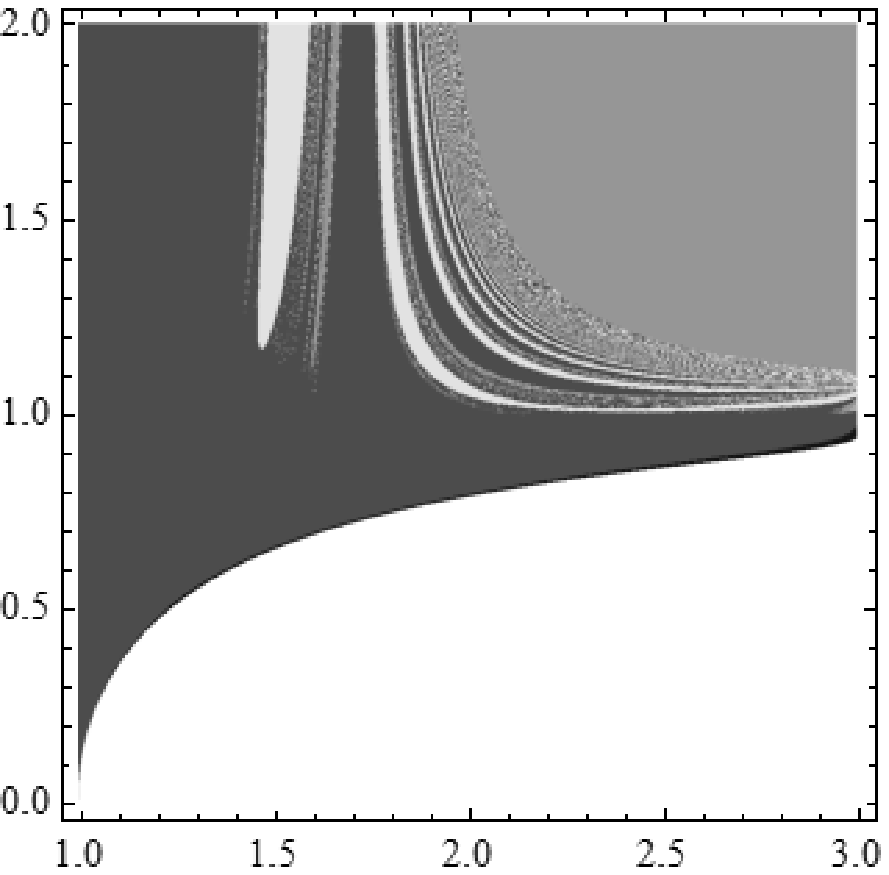}
\hspace{1cm}\includegraphics[width=7cm]{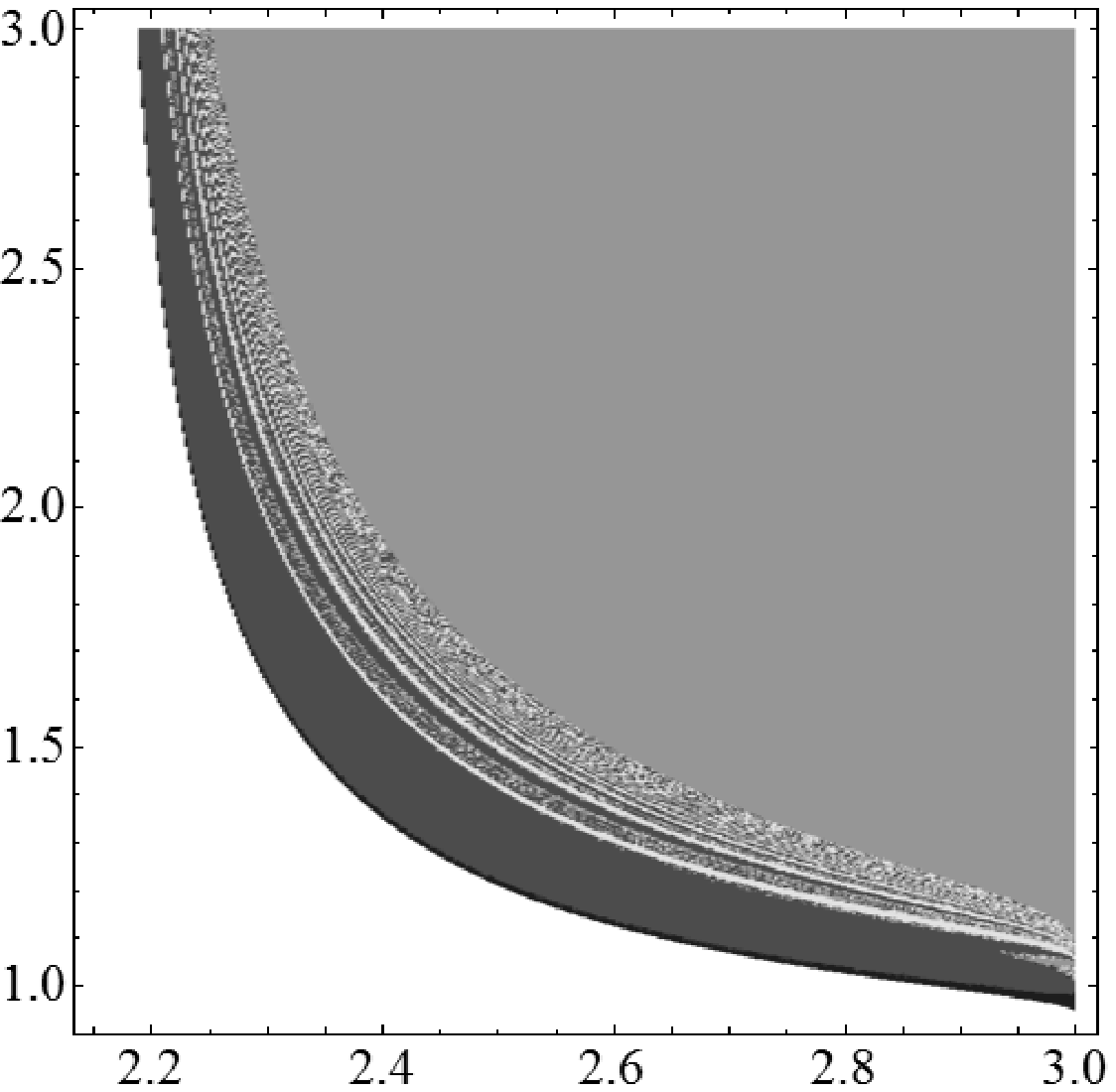}\non\\
&&\hspace{4.0cm}({\bf a})\hspace{7.8cm}({\bf b})\non
\ea
\caption{Basin-boundary plots for a charged particle kicked from the ISCO of the given $\rho_{o}$ (horizontal axis) defined by the magnetic field $b$ with different energies (vertical axis). The dark grey zones correspond to capture, the grey zones correspond to escape to $z\to+\infty$, while the light grey zones correspond to escape to $z\to-\infty$. Plot ({\bf a}): $\ell>0$ and plot ({\bf b}) $\ell<0$. The step size for both $\rho$ and $\ce$ is $2.5\times 10^{-3}$.}\label{F3}
\end{center}
\end{figure*}
\begin{figure*}[ht]
\begin{center}
\ba
&&\hspace{1cm}\includegraphics[width=7cm]{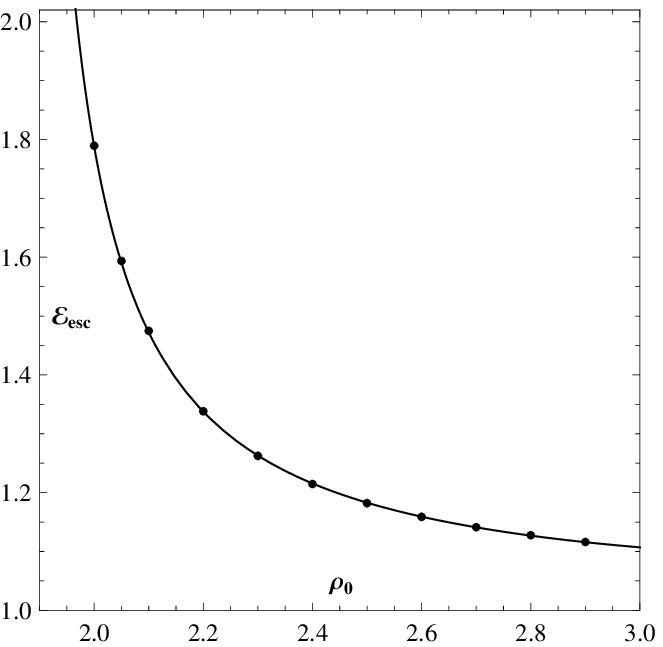}
\hspace{1cm}\includegraphics[width=7cm]{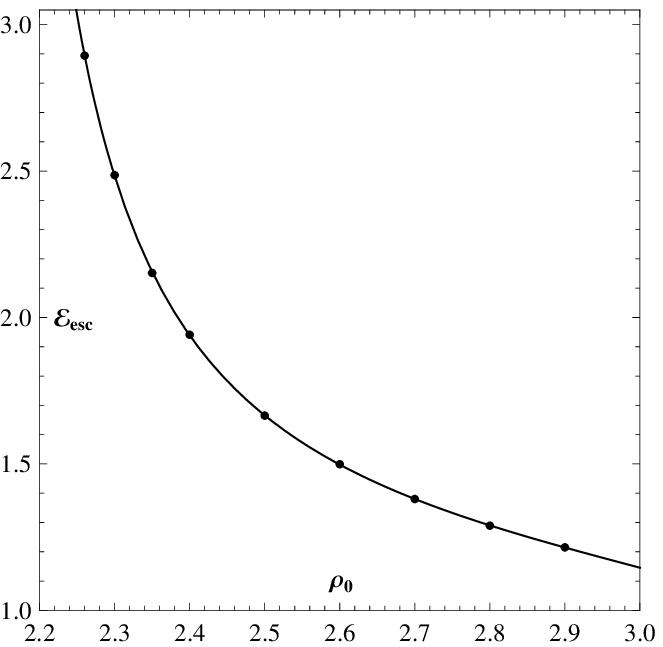}\non\\
&&\hspace{4.0cm}({\bf a})\hspace{7.8cm}({\bf b})\non
\ea
\caption{Critical escape energy for $\ell>0$ [plot ({\bf a})] and $\ell<0$ [plot ({\bf b})]. The curves are given by the analytical expression \eq{Ecr1} for plot ({\bf a}) and by \eq{Ecr2} for plot ({\bf b}).}\label{F5}
\end{center}
\end{figure*}
\begin{figure*}[ht]
\begin{center}
\ba
&&\hspace{1cm}\includegraphics[width=6.8cm]{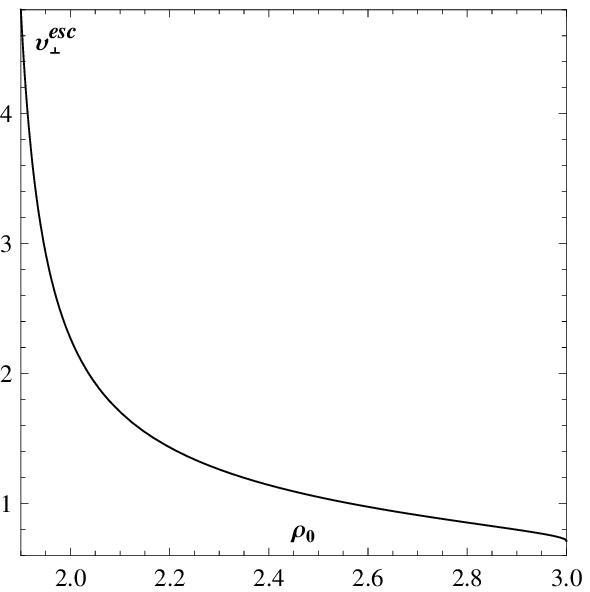}
\hspace{1cm}\includegraphics[width=6.8cm]{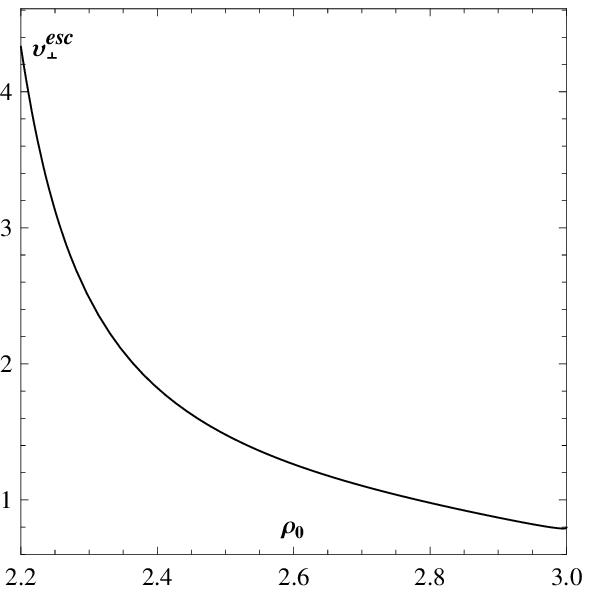}\non\\
&&\hspace{4.0cm}({\bf a})\hspace{7.8cm}({\bf b})\non
\ea
\caption{Critical escape velocity for $\ell>0$ [plot ({\bf a})] and for $\ell<0$ [plot ({\bf b})].}\label{F5a}
\end{center}
\end{figure*}
\begin{figure}
        \begin{subfigure}[b]{0.5\textwidth}
                \centering
                \includegraphics[width=8.0cm]{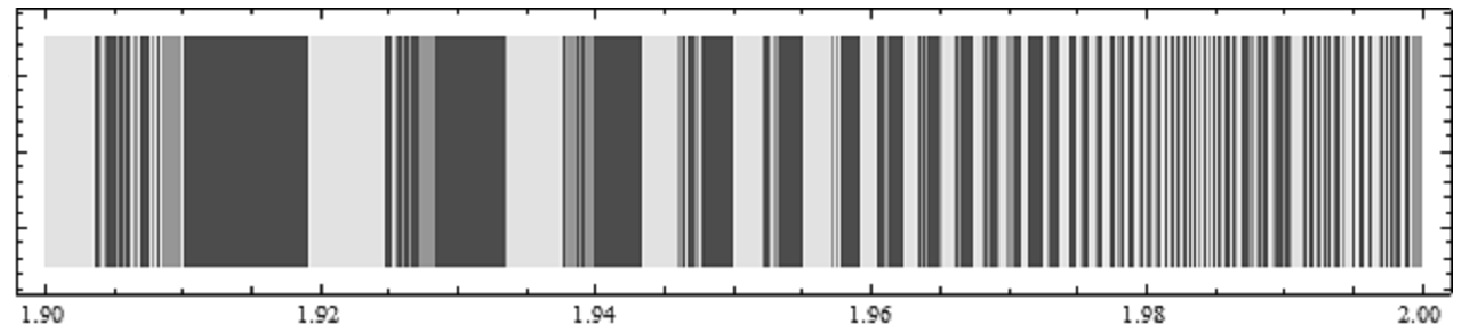}
                \caption{A stripe at $\ce$ around $1.9$ for $\ell>0$.}
        \end{subfigure}
        \begin{subfigure}[b]{0.5\textwidth}
                \centering
                \includegraphics[width=8.0cm]{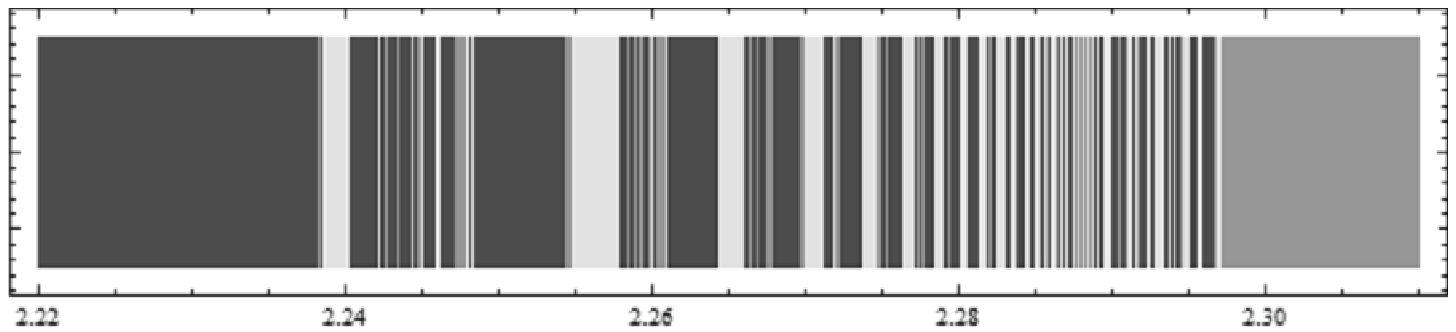}
                \caption{A stripe at $\ce$ around $2.5$ for $\ell<0$.}
        \end{subfigure}
        \caption{Magnified stripes from the fractal region in the basin-boundary plots.}
        \label{F4}
\end{figure}

To distinguish the final states  of the particle we introduce an integer number $n$. It takes the value 0 for the capture case and $\pm 1$ for escape to $z\to\pm\infty$, respectively.

If one focuses on a single trajectory, one can see that there exists a variety of its types with qualitatively different behavior of the particle within the domain close to the black hole. For each of these types of the motion the particle may pass a number of times through the equatorial plane. To characterize the dynamical behaviour of the particle in this domain we introduce the ``winding'' number $w$ counting how many times the particle crosses the equatorial plane before it gets captured or escapes to the spatial infinity. The integer number $w$ is, evidently, a topological invariant. For our choice of the initial conditions, the particle after the kick starts its motion from the equatorial plane in the positive $z-$direction. For this reason, it is evident, that for $n=+1$ the winding number $w$ is even, while for $n=-1$ it is odd. In particular, when a particle goes to $z\to+\infty$ without further crossing the equatorial plane, $w=0$.

In rare cases of the computations the particle stays in the vicinity of the equatorial plane during the numerical lifetime $\sigma$. It crosses the plane many times and forms a compact cloud in the corresponding phase space. However, we expect that such a particle eventually falls into the black hole or escapes to the infinity.

Figures~\ref{F1a}--\ref{F1d} illustrate possible ``capture" and ``escape" trajectories of the charged particle near the magnetized black hole for both $\ell>0$ at $\rho_{o}=2$ and $\ell<0$ at $\rho_{o}=2.5$ cases. They collect several examples of such trajectories for different values of the invariants $n$ and $w$.

The fate of a kicked particle was found to be extremely sensitive to the initial conditions: even a very tiny change of these conditions may drastically modify its global behavior. Such an extreme sensitivity is an indication of non-integrability of the system and its chaotic nature. The dynamical system \eq{eq1} is Hamiltonian and conservative. Chaos in Hamiltonian systems was extensively studied (see, e.g., \cite{ott} and the references therein). However, most of the known tools and results, such as construction of the Poincar\'e sections and the KAM theorem, are related to (quasi) periodic bounded type of motion, while in our case the particle's motion can be unbounded. For an analysis of our dynamical system we shall use the methods developed for the study of chaotic dynamical systems which posses several attractors. In our case, we have three different types of asymptotic trajectories. For such a  motion a phenomenon of fractal basin boundaries takes place. We shall discuss it in the next section.

\section{Basin-boundary analysis}

\subsection{Basin-boundary plots}

The chaotic nature of a dynamical system is exhibited in its sensitive dependence of trajectories on initial conditions. A special case is, what is called, the {\em final state sensitivity}. Such a sensitivity takes place if a dynamical system has several coexisting attractors.  As a result, for given initial conditions, a dynamical trajectory will typically converge to one of the attractors. Therefore, there must be boundaries of basins of the attractors  separating the sets of trajectories with different final states. Such boundaries are often fractals. The larger is the fractal dimension of the basin boundaries the less predictable is the behavior of the dynamical system. To measure fractal dimension one can use the box-counting dimension. Details and additional information can be found in, e.g. \cite{chaos,ott}. This method, in particular, was used in the analysis of another dynamical system appearing in General Relativity (see, e.g. \cite{FrLa}).

\begin{figure*}[ht]
\begin{center}
\ba
&&\hspace{1cm}\includegraphics[width=7cm]{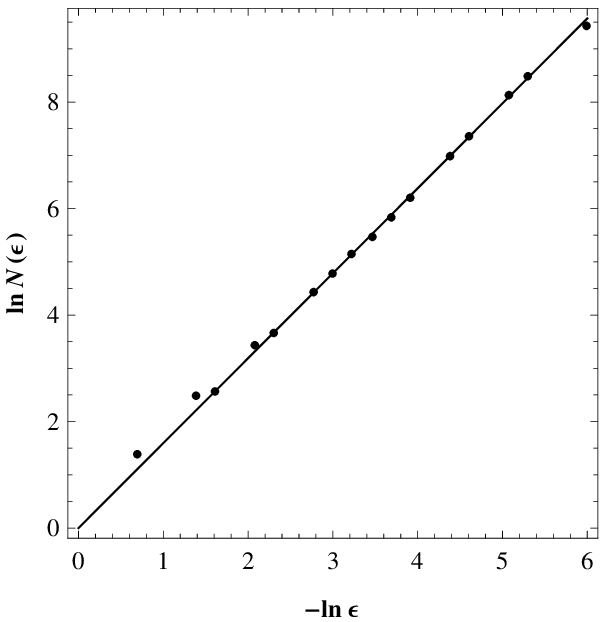}
\hspace{1cm}\includegraphics[width=7cm]{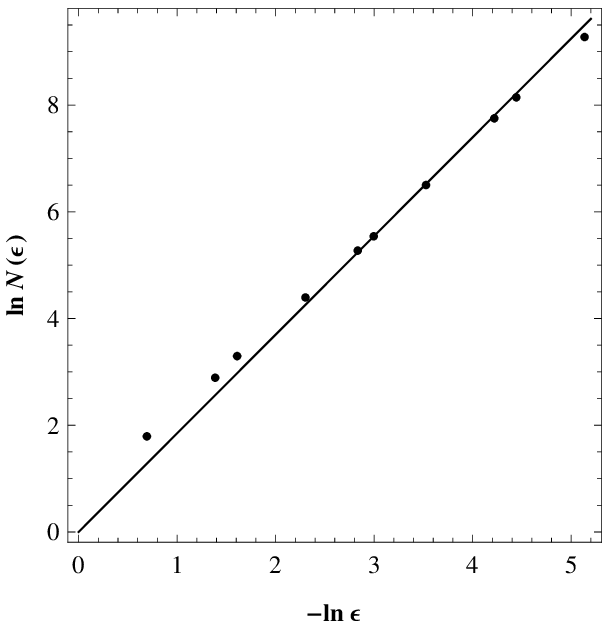}\non\\
&&\hspace{4.0cm}({\bf a})\hspace{7.8cm}({\bf b})\non
\ea
\caption{The box-counting dimension. Plot ({\bf a}): $\ln\,N(\epsilon)$ vs. $\ln(1/\epsilon)$ for $\ell >0$. Plot ({\bf b}): $\ln\,N(\epsilon)$ vs. $\ln(1/\epsilon)$ for $\ell <0$.}\label{F6}
\end{center}
\end{figure*}

Here we use the basin-boundary method for the  analysis of the asymptotic behavior of the particle trajectories. The corresponding plots are presented in Fig.~\ref{F3}. As we mentioned, the particle is initially at the ISCO. Plot ({\bf a}) corresponds to the  case $\ell>0$ and plot ({\bf b}) to the case $\ell<0$. The horizontal axis on these plots shows the dimensional parameter $\rho_{o}$ of the ISCO, while the vertical axis shows the dimensionless energy $\ce$ of the particle after the kick. The white domains in these plots correspond to the region forbidden for the particle's ISCO. The shadowed regions in these plots consist of square pixels of the side size $\rho$ and $\ce$ equal to $2.5\times 10^{-3}$. We used three different colors for these pixels. The color of a pixel determines the final outcome of the particle motion.
These colors are chosen so that dark grey corresponds to the particle capture, grey color corresponds to escape to $z\to+\infty$ ($n=1$), while light grey corresponds to escape to $z\to-\infty$ ($n=-1$).

\subsection{Critical escape energy and velocity}

The general structure of the plots shown in Fig.~\ref{F3} can be described as follows:
The dark grey region which adjoints to the white region at the bottom of the plots corresponds to the particle capture. The uniformly grey region in the upper-right part of the plots corresponds to the particle escape to $z\to+\infty$. The winding number in this region is $w=0$. This region is restricted from below by a diffuse domain. The uniform region is separated from the diffuse domain by the line which we call the critical escape energy line. Using our numerical results we can estimate the critical escape energy of the particle describing this line. The approximate analytical expression for the case $\ell>0$ is
\be\n{Ecr1}
\ce_{\text{esc}}\approx1+\frac{0.115(3.463-\rho_{o})}{(\rho_{o}-1.851)(3.433-\rho_{o})}\,.
\ee
The maximal relative error of the escape energy is $0.2\%$. Plot \ref{F5}({\bf a}) illustrates the escape energy $\ce_{\text{esc}}$. We can obtain a similar approximate analytical relation for the escape energy for the $\ell<0$ case. It reads
\be\n{Ecr2}
\ce_{\text{esc}}\approx1+\frac{0.4393(3.198-\rho_{o})}{(\rho_{o}-2.105)(3.667-\rho_{o})}\,,
\ee
where the maximal relative error is also $0.2\%$. This expression is illustrated in plot \ref{F5}({\bf b}).
The critical escape velocity $\vp^{esc}$ as a function of $\rho_{o}$ can be derived using these expressions together with Eqs. \eq{e0}--\eq{lb}. Plots ({\bf a}) and ({\bf b}) in Fig.~\ref{F5a} illustrate the critical escape velocity for $\ell>0$ and $\ell<0$, respectively.

\subsection{Near-critical behavior}

For a given $\rho_o$ and energies close but less than the critical one the final state of the particle cannot be strictly predicted. The corresponding near-critical domain contains the final states of all the three different types. To illustrate it more clearly let us consider a small stripe in these domains. Magnifications of the stripes are shown in Fig.~\ref{F4}. The horizontal axis on these plots shows the dimensional parameter $\rho$ of the ISCO, while the vertical axis shows the dimensionless energy $\ce$ of the particle after the kick which chosen close to $1.9$ for $\ell>0$ and close to $2.5$ for $\ell<0$. These magnified plots demonstrate a linear structure of different regions corresponding to capture and both the types of the escape corresponding to $n=\pm1$. Similar magnified stripes can be constructed for different values of the magnification factor. The remarkable fact is that each of such plots has similar structure which does not depend on the value of the magnification. In other words, the near-critical diffuse domain has fractal structure.  This fractal structure is a very complicated Cantor-set like structure, such that a magnification of any portion of the fractal region reveals similar pattern of the escape and capture regions on a smaller scale and it continues ad infinitum. In the fractal regions the winding number $w$ corresponding to either escape or capture can take different values and generally increases with the increasing repetition of the patterns.

\subsection{Fractal dimension of the near-critical domains}

To get a qualitative measure of the complexity of the fractal regions we calculate the box-counting fractal dimension $D_f$,
\be
D_f\equiv\lim_{\epsilon\rightarrow 0}\frac{\ln{N(\epsilon)}}{\ln{\frac{1}{\epsilon}}}\hh 1\leq D_f<2\,,
\ee
where $N(\epsilon)$ is the number of squares of the side-length $\epsilon$ needed to cover a basin boundary. Such squares are counted only if they contain at least two different colors. The box-counting fractal dimension gives us a quantitative measure of uncertainty in our numerical computations (see, e.g., \cite{ott, chaos}). Namely, if our current uncertainty is, say $\Delta$, and we want to reduce it by a factor of $10^{-n}$ by improving the precision of our computations, then the necessary precision of our computations, say $p$, is given by
\be
\log_{10} p=\log_{10} \Delta -\frac{n}{\alpha}\,,
\ee
where
\be
\alpha=2-D_f\hh 0<\alpha\leq 1
\ee
is the uncertainty exponent. Thus, we need $n/\alpha$ additional digits to achieve the desired precision.

Figure \ref{F6} contains plots of $\ln{N(\epsilon)}$ vs. $\ln{(1/\epsilon)}$ for different values of $\epsilon$ for the fractal structures shown in Fig.~\ref{F3}. The plots illustrate a linear relation for sufficiently small $\epsilon$. The fractal dimensions of the two basins-boundaries are
\ba
D&\approx&1.60  \hh \ell>0, \\
D&\approx&1.85  \hh \ell<0.
\ea
The fractal dimension is closer to 2 for the $\ell<0$ case. Thus, the uncertainty exponent $\alpha\approx 0.15$ is smaller than $\alpha\approx 0.40$ for the case of $\ell>0$, and the corresponding near-critical domain has more complex fractal structure. In this case an increase in precision requires more digits in computations.

\subsection{Additional details of the basin-boundary plots.}

We described the main features of the structure and domains in the basin-boundary plots. However, these plots contain additional structure which we briefly describe now. First of all, let us mention that for $\ell>0$ and the values of $\rho_o$ in the vicinity of $\approx 1.5$ and $\ce\gtrapprox 1.2$ there is an escape lagoon illustrated by the light-grey color. For the initial data corresponding to this lagoon the charged particle also escapes to infinity but in the direction opposite to the initial kick, that is with $n=-1$. The winding number in this region is $w=1$. Besides this, there are also smaller size light grey regions which correspond to different values of the winding number. These regions form a well-visible set of the light grey stripes located to the right of the lagoon. Similar light grey stripes corresponding to the backscattering to $z\to-\infty$ are present for $\ell<0$ but they are much less profound.

\section{Summary}

We studied a charged particle motion in the spacetime of a weakly magnetized black hole. We demonstrated that the  space of its trajectories has rather a rich structure. There exist three different types of asymptotic behavior: capture and escape to the asymptotic spatial infinity, $z\to\pm\infty$. There also exists  a class of escape trajectories when the charged particle spends a considerable time moving in the vicinity of the black hole close to its equatorial plane, crossing it again and again. For such a particle its escape to infinity has features similar to a diffusion process.

Certainly, our model is rather simplified. We made several assumptions that simplify the problem. These simplifications are of two different types. First, we chose a special type of the magnetic field. In a `realistic' case the magnetic field decreases at far distances. The other simplification was the choice of  orbits and parameters of the `kicks'.  It should be emphasized that the basin-boundary plots were constructed for a very special case of the kicking mechanism. For other more general types of kicks the corresponding basin-boundary plots might be modified. However, we expect that the following three main features are common: Namely, the motion of the kicked particle is mainly chaotic. There is a critical escape energy line (surface) and the near-critical domains which have fractal structure. It is interesting to confirm this by direct numerical calculations. Another interesting generalization of the problem is an analysis of the critical escape phenomenon and structure of near-critical domains for rotating black holes in the presence of a magnetic field.

\begin{acknowledgements}

The authors (V.F and A.S) are grateful to  the Natural Sciences and Engineering Research Council of Canada for the financial support. The author (V.F.)  thanks also  the Killam Trust for its support.

 \end{acknowledgements}

\end{document}